\documentclass[preprint,12pt]{elsarticle}

\usepackage{amssymb}
\usepackage[modulo]{lineno}

\usepackage{xspace}
\usepackage{fancyvrb}
\usepackage{fvextra}
\usepackage{relsize}
\usepackage{overpic}
\usepackage{listings}
\usepackage{hyperref}
\usepackage{cleveref}
\usepackage{color}
\usepackage{cancel}
\usepackage{soul}

\newcommand{\geant}{\textsc{Geant4}\xspace}
\newcommand{\allpix}{Allpix$^2$\xspace}
\newcommand{\allpixEFF}{Allpix$^2$+EFF\xspace}
\newcommand{\allpixTCAD}{Allpix$^2$+TCAD\xspace}

\newcommand{\um}{\ensuremath{\mu}m\xspace}
\newcommand{\Emag}{\ensuremath{\sqrt{\Sigma E_i^2}}\xspace}
\newcommand{\IntEdl}{\ensuremath{-\int{\vec{E}\cdot{\rm d}\vec{\ell}}}\xspace}

\journal{Nucl. Instr. Meth. A}

\begin{document}

\begin{frontmatter}

\title{\boldmath A derivation of the electric field inside MAPS detectors from beam-test data and limited TCAD simulations}

\author[a]{A. Santra\fnref{f1}} 
\author[a]{N. Tal Hod\fnref{f2}}
\fntext[f1]{arka.santra@weizmann.ac.il}
\fntext[f2]{noam.hod@weizmann.ac.il}

%\affiliation[a]{
%organization={Department of Particle Physics and Astrophysics,Weizmann Institute of Science},
%city={Rehovot},
%postcode={7610001}, 
%country={Israel}
%}

\address[a]{Department of Particle Physics and Astrophysics,Weizmann Institute of Science, Rehovot, 7610001, Israel.}

\begin{abstract}
Solid semiconductor sensors are used as detectors in high-energy physics experiments, in medical applications, in space missions and elsewhere. Minimal knowledge of the electric field inside the elementary cells of these sensors is highly important for their performance understanding. The field governs the charge propagation processes and ultimately determines the size and quality of the electronic signal of the cell. Hence, the simulation of these sensors as detectors in different analyses relies strongly on the field knowledge. For a certain voltage applied to the cell, the field depends on the specifics of the device's growth and fabrication. The information about these is often commercially protected or otherwise very difficult to encode in state-of-the-art technology computer-aided-design (TCAD) software. In this work, we show that by taking the top-down approach, combining public beam-test data and a very limited public TCAD knowledge, we are able to effectively approximate the 3D electric field function in the pixel cell of one important and widely used example, namely the ALPIDE sensor, for simulating the charge propagation processes. Despite its broad usage worldwide, the ALPIDE field is not available to the community. We provide an effective field function, that adequately describes the sensor behaviour without trying to reconstruct further details about the device or the details behind its processing. We comment on the process by which the effective field function is derived with the help of the \allpix software, and on how similar work can be performed for other devices, starting from the same grounds.
\end{abstract}


\end{frontmatter}

\section{Introduction}
\label{sec:intro}
Solid semiconductor sensors are used for imaging, tracking and calorimetry in different disciplines.
When an ionizing particle traverses the elementary cell of the sensor (pixel, strip, pad, etc.), an electric current pulse is produced at the collection node of that cell (the electronic signal).
This current is in effect produced from electrons or holes (charge carriers), which are created in pairs along the trajectory of the ionizing traversing particle.
The collected charges are processed by the front-end electronics and a hit is formed if the collected charge is above a certain threshold.
The number of charge carriers and the way these propagate to the collection node depend primarily on the sensor material.
The electronic signal depends in turn strongly on the material, but it also depends on a number of other parameters such as the sensor growth and fabrication processes, the sensor geometry, the external voltage applied to the sensor, the damage the sensor may have suffered due to continuous exposure to radiation, etc.

A key ingredient in the understanding of the charge carriers' propagation in the cell is the electric field inside it.
The field governs this propagation process and ultimately determines the size and quality of the electronic signal output of the cell.
Hence, the simulation of these sensors relies strongly on the electric field knowledge.
When designing a new sensor,
an important task in the process is the implementation of the corresponding technology computer-aided-design (TCAD) simulation of the device\footnote{This is usually done using Sentaurus software from SYNOPSYS~\cite{SYNOPSIS} or ATLAS software from SILVACO~\cite{SILVACO}.}.
One of the outputs of such TCAD simulation is the fine-grained 3D electric field map inside the cell.
Even from the manufacturer's side, implementing the device's specifics in TCAD is an extremely complicated task.

The TCAD simulation is particularly important for new monolithic active pixel sensors (MAPS)~\cite{SNOEYS2013125}, which are used more and more extensively in recent years thanks to the development of sub-micron capabilities in different foundries worldwide.
MAPS are now becoming popular, e.g., for the next-generation high-energy physics experiments (for tracking, vertexing and calorimetry) and elsewhere.
It may happen that users of a specific sensor do not have access to the TCAD field map of the non-trivial field.
This poses a serious problem for these users, who effectively lack the first building block of the device simulation and therefore cannot properly simulate its response as a detector.
To overcome this limitation, we show that combining public beam-test data and a very limited public TCAD simulation output knowledge, we are able to effectively approximate the 3D electric field function in the pixel cell of one important and widely used example: the ALPIDE monolithic silicon sensor~\cite{ALPIDE1,ALPIDE2,ALPIDE3,ALPIDE4,ALPIDE5}.
The ALPIDE (ALice PIxel DEtector) sensor was developed by and for the ALICE~\cite{ALICE} experiment at the LHC together with TowerJazz~\cite{TJ}.
The inner tracking system (ITS) of ALICE~\cite{ALICEITS} has been upgraded successfully in 2021 with this new technology and it is now the largest pixel detector ever built.
The successful ALPIDE technology is also being used elsewhere in other experiments/facilities like LUXE~\cite{LUXECDR}, sPHENIX~\cite{PHENIX:2015siv,Dean:2021rlo}, DESY's beam-test facility~\cite{DESYTELESCOPE}, the CSES-2 satellite mission~\cite{Iuppa:2021ozs} in space, Proton Computed Tomography (pCT)~\cite{pCT1,pCT2}, and more.
However, despite its wide usage, the TCAD electric field map is not available to the community as it is under proprietary restriction.

In this work, we discuss how the 3D effective field function (denoted EFF hereafter) in the INVESTIGATOR pixel sensor can be adequately approximated.
This is done in a ``top-down'' approach rather than in a ``bottom-up'' approach, where one would need to start from first principles (encoding the doping profiles, Maxwell's equations, etc.) as in TCAD.
The INVESTIGATOR is a very close variation of the ALPIDE sensor's production-version, sharing the same wafer and collection diode characteristics.
Specifically, the diode geometry and the epitaxial layer thickness are the same (25~\um), while the sensitive volume is slightly different ($28\times 28~\mu$m$^2$ vs $29.24\times 26.88~\mu$m$^2$)~\cite{MIKOPAPER}.
Most importantly, the ALPIDE sensor is digital while the INVESTIGATOR sensor has analog output, allowing to read out the charge.
There are further smaller differences but these can be regarded as sufficiently small for the purpose of this work.
Therefore, we assume that a simple re-scaling of the pixel dimensions is enough to transform the EFF from the INVESTIGATOR geometry to the ALPIDE geometry, similarly to what is done in~\cite{MIKOPAPER}.
The way the EFF is formulated, with respect to the pixel dimensions in 3D, allows to easily do this simple transformation.

We use the \allpix software~\cite{allpix,allpixmanual} to compare the performance in simulation between two cases, once starting from our 3D EFF and once starting from public results that are using the actual TCAD field map, which remains unknown to us.
The two cases are denoted hereafter as \allpixEFF and \allpixTCAD, respectively.
For the comparison, we use the \allpixTCAD results shown in~\cite{DANNHEIM2020163784}.
The authors of~\cite{DANNHEIM2020163784} go further to compare their \allpixTCAD simulation with data collected in a beam-test~\cite{TESTBEAM}.
All results in~\cite{DANNHEIM2020163784}, \cite{TESTBEAM} and here are obtained using the INVESTIGATOR sensor.

The performance of the three scenarios (\allpixEFF, \allpixTCAD and data) is shown to be very similar.
This agreement gives confidence in our EFF as the cornerstone for further simulation campaigns, where the INVESTIGATOR (and ALPIDE) sensor is used.
We comment on the derivation process and provide the EFF as a set of ROOT~\cite{rene_brun_2019_3895860} TFormula-compatible strings in~\cite{gitEField}.
We also provide the code that produces our EFF for further fine-tuning and such that similar work can be started for other sensors.
We also provide the \allpix configuration files used in our simulation.

\section{The \texorpdfstring{\allpix}{} setup}
\label{sec:allpix}
The \allpix software~\cite{allpix,allpixmanual} simulates the processes triggered inside different semiconductor devices, when ionizing particles traverse these.
Particularly, it simulates the electron-hole pair creation, the drift or diffusion of these charge carriers given some electric field, the charge collection by the electrodes and finally the digitization of collected charge along with the electronic noise in the front-end electronics.
The task of simulating the energy deposition process by the ionizing particles (an input for the charge carriers generation) is  external to \allpix.
Usually this is done by \geant~\cite{geant1, geant2, geant3}, which is interfaced directly to \allpix.
The output of the simulation chain contains the fired pixels, where the collected charge passes some predefined threshold.
An induced charge generation in adjacent pixels (to the pixel where the ionizing particle passes through) is also simulated by \allpix.
The list of fired pixels can then be used in a subsequent analysis.

The \allpix simulation configuration for this work is kept identical to that of~\cite{DANNHEIM2020163784} except for the electric field input.
This is necessary for the validation of the EFF by comparing the \allpixEFF and \allpixTCAD performances.
Particularly, the total thickness of the sensor in our \allpixEFF simulation samples is taken to be 100~\um, with an epitaxial layer thickness of 25~\um and a pitch of $28\times 28~\mu$m$^2$ in $x\times y$, like the INVESTIGATOR sensor used in~\cite{DANNHEIM2020163784}.
While we configure our code that generates the 3D EFF to have the same parameters as for the INVESTIGATOR sensor, we make sure that it is easy to reconfigure these parameters using the production version of the ALPIDE.
For example, the pixel pitch is defined as a global parameter in the code and hence, it can be simply re-set at one place such that the EFF would scale naturally.
The bias voltage applied to the INVESTIGATOR p-well in~\cite{DANNHEIM2020163784} and~\cite{TESTBEAM} is $-6$~V and a voltage of $+0.8$~V is applied to the collection electrode.
When using the TCAD field map, the voltage settings are dictated by that.
In the user-defined function case, however, the settings are completely set by the function's normalization.
This normalization can be capped roughly by $\sim -6$~V, when integrating the EFF along the epitaxial layer, but it strongly depends on the shape as well.

\section{The process of deriving the effective field function}
\label{sec:reveng}
The detailed TCAD map of the electric field inside the INVESTIGATOR (and ALPIDE) sensor is unavailable publicly.
However, its magnitude, \Emag (with $i=x,y,z$), in the $x-z$ plane for $y=0$ is shown in figure~4 of~\cite{DANNHEIM2020163784}.
The field lines in this plane are also overlaid in this figure.
The magnitude of the field at the faces of the pixel in three-dimensions is shown in figure~3 of~\cite{DANNHEIM2020163784}.
Due to the various restrictions, these two figures are given without axes scale-labels.
The meaning of the colors in these figures can be loosely interpreted from the field of a similar sensor shown in figures~8.4 and 8.6 of an older work from 2018~\cite{magdalena}.
This work used the CLIC Tracker Detector (CLICTD), a monolithic chip~\cite{Kremastiotis:2020msg} which is also produced in a 180~nm imaging CMOS process on a high-resistivity epitaxial layer like the INVESTIGATOR (and ALPIDE).
The extensive TCAD studies in~\cite{magdalena} show a similar field behavior to the one of the INVESTIGATOR, where the 2D shape of the field's components are given, including the transverse ones.
This information allows to understand the different symmetries of the shape better and it is also used to define our EFF.

The field's magnitude shape is highly non-linear and the EFF cannot be easily deciphered from the available information.
At this step, we therefore only have a rough idea on (i) how the magnitude of the 3D EFF should look like in a few slices of the pixel and (ii) how the $x$ and $y$ components of the field behave in a few slices of the pixel.
We also know that besides mimicking the shape visually, its magnitudes' normalizations should result in a voltage of $\sim -6$~V, when integrated along $z$ in the upper 25~\um of the sensor.

This information is clearly not enough to approximate the field effectively.
However, we do have another few indirect pieces of invaluable information from~\cite{DANNHEIM2020163784} detailing the performance of the sensor with the 120~GeV $\pi^+$ particle beam of~\cite{TESTBEAM}.
This includes the charge distributions (figure~9), the cluster size distributions (figures~10-11) and the position residual distributions in $x$ (figure~15), all given for a specific threshold.
The behavior for different thresholds can be seen in the cluster size graphs (figure~13), the spatial resolution graphs in $x$ and $y$ (figure~16) and the efficiency graphs (figure~18).
With this, we can iteratively plug in one EFF ansatz at a time such that
\begin{enumerate}
\item its 3D magnitude, \Emag (with $E_i=f_i(x,y,z)$ and $i=x,y,z$), visually resembles the magnitude of the TCAD field shape in the slices, including the field lines from figures~3-4 of~\cite{DANNHEIM2020163784} (and figure 8.4 from~\cite{magdalena}),
\item its transverse components follow the same symmetry as shown in figure~8.6 of~\cite{magdalena}, noting the sign flip between positive and negative $x$,$y$,
\item it results in an integral $\IntEdl \simeq -6$~V, where $\vec{\ell}$ runs along the negative $z$-axis at the top 25~\um of the sensor in its center, and
\item it gives a good agreement with the performance figures~9-18 of~\cite{DANNHEIM2020163784} after simulating the same scenarios with our \allpixEFF setup. The criteria for a good agreement are quantified below.
\end{enumerate}

For steps~1 and~2 above, we add the features of the field, one by one and plot the field magnitude in 3D and a few 2D slices, similar to the ones available from TCAD in~\cite{DANNHEIM2020163784}.
This is done initially with simple functional shapes (sphere, arcs, stripes, Gaussian and exponential shapes, etc.).
Upon adding a new feature, we verify that it merges properly with the existing features in terms of the relative normalization and the smoothness in the transition regions.
In case the field lines are also available from TCAD in some 2D slices, they must always match in direction.
This is adjusted by changing both the feature's relative normalization and their sign in all three components.
When adding transverse components, a parity transformation between positive and negative $x$ and $y$ must be respected (see figure~8.6 of~\cite{magdalena}).
The $z$ component must have a rotational symmetry around the center of the pixel (see figure~8.4 of~\cite{magdalena}).

For step~4 above, we initially simulate 1000 primary $\pi^+$ beam particles (at 120~GeV) to see if the cluster charge and size distributions roughly agree with those in figures~9-10 of~\cite{DANNHEIM2020163784} for a nominal threshold of 120~e.
If these agree, we continue to check a wider range of thresholds (40-700~e) with larger statistics (20,000 primaries) and check the agreement with figures~9-18 of~\cite{DANNHEIM2020163784}.
For the comparison, we digitize the figures from~\cite{DANNHEIM2020163784} using the WebPlotDigitizer software~\cite{DIGITIZEPLOTS}.
The uncertainties on the \allpixTCAD shapes are taken from the ratio panels wherever these are available.
Otherwise, they are taken from the distributions themselves.

A good agreement is judged according to several criteria, depending on the distributions or graphs from~\cite{DANNHEIM2020163784}.
We note that each point in the graphs is a summary of one distribution (for example, the resolution vs threshold).
For distributions, we require that the ratio between our \allpixEFF and the \allpixTCAD simulation is within 10\% of unity across the regions where the statistics are high, taking into account the uncertainties.
The same is required for the graphs.
We also define the pull,
\begin{equation}
{\rm Pull}=\frac{\rm E-T}{\sqrt{({\rm D-T})^2+\delta^2_{\rm E} + \delta^2_{\rm T}}},
\label{eq:pull}
\end{equation}
where E stands for the distribution coming from the \allpixEFF, T stands for \allpixTCAD and D stands for the Data.
The quadrature in the denominator contains the statistical error on E, the uncertainty on T and the difference between D and T.
For the distributions, the pull from equation~\ref{eq:pull} takes a binned form and we also quote its weighted average (using the bin height as a weight) in the full range.
For graphs, we show the respective ratio and pull from equation~\ref{eq:pull} for each point.
The requirement for a good agreement between two distributions is that the binned pull distribution is within $\pm 3$ in the range where the statistics is high and the weighted average of the pull is $<1$ across the full range.
The requirement for a good agreement between two graphs is that the pulls are within $\pm 3$.
We interpolate the data graphs to calculate the pull for simulated points between the (fewer) available data points.
As another measure of comparison, we check the cluster charge distribution's most-probable-value (MPV) between the different distributions, requiring that the results are compatible within $<10\%$.

The iterative process of adding field features to the EFF ansatz and testing the performance (per ansatz) may be stopped when all four criteria above are satisfied.
The requirements listed above should depend on the specific application for which the EFF is to be used.
Hence, one can define tighter/looser requirements and/or limit the range where these are tested.
Specifically for the INVESTIGATOR sensor, about a hundred such iterations were needed in order to arrive at a satisfactory EFF in the range of thresholds interesting for normal operation of the sensors in experiments (above $\sim 100$~e).
The time of processing for each such iteration can be relatively short with $\mathcal{O}(<1~{\rm h})$ per iteration, depending on the availability of high performance computing cluster.
Therefore, the overall convergence time depends on the fine-tuning of the EFF expression itself.

\section{The effective field function of the INVESTIGATOR sensor}
\label{sec:elfieldalpide}
\begin{figure}[!ht]
\centering
\begin{overpic}[width=0.465\textwidth]{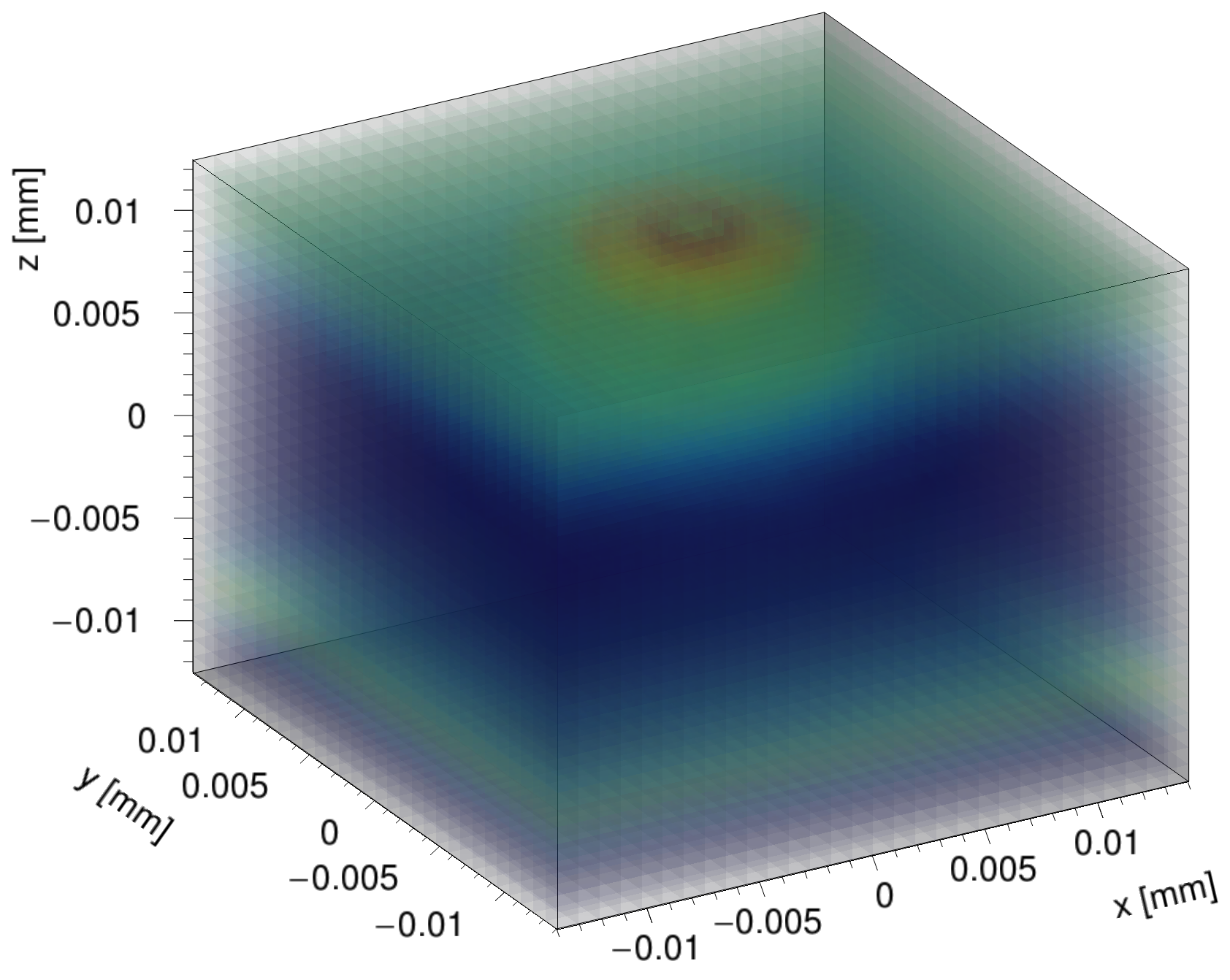}\end{overpic}
\begin{overpic}[width=0.525\textwidth]{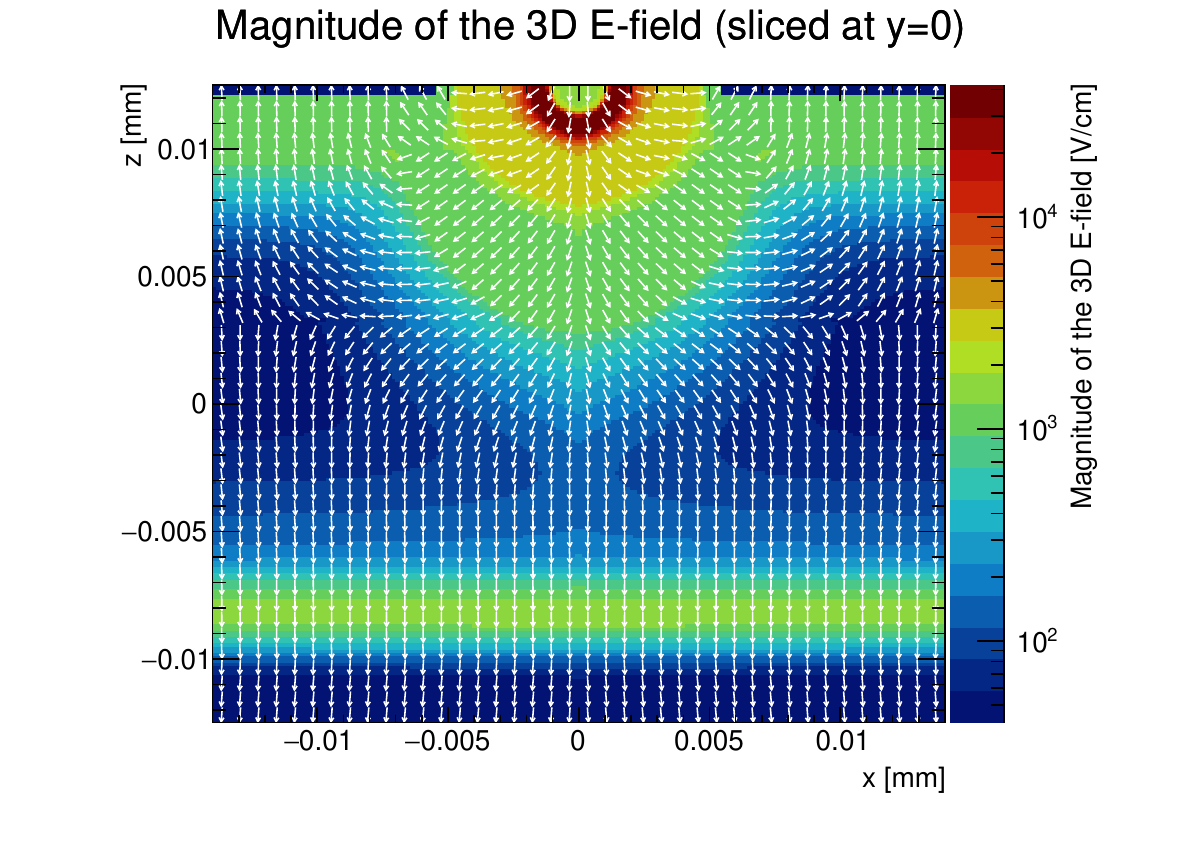}\end{overpic}
\caption{Left: the 3D EFF magnitude at the INVESTIGATOR pixel sensor faces (sides and top). Right: the EFF magnitude in the $z$ vs $x$ plane sliced at $y=0$ overlaid with the field lines shown as white arrows (the arrows are positioned at the bin centers). The shapes are given for the upper 25~\um of the sensor, where the EFF is non-zero. These plots should be compared with figures~4 and~3 of~\cite{DANNHEIM2020163784}, respectively.}
\label{fig:effmag1}
\end{figure}

\begin{figure}[!ht]
\centering
\begin{overpic}[width=0.99\textwidth]{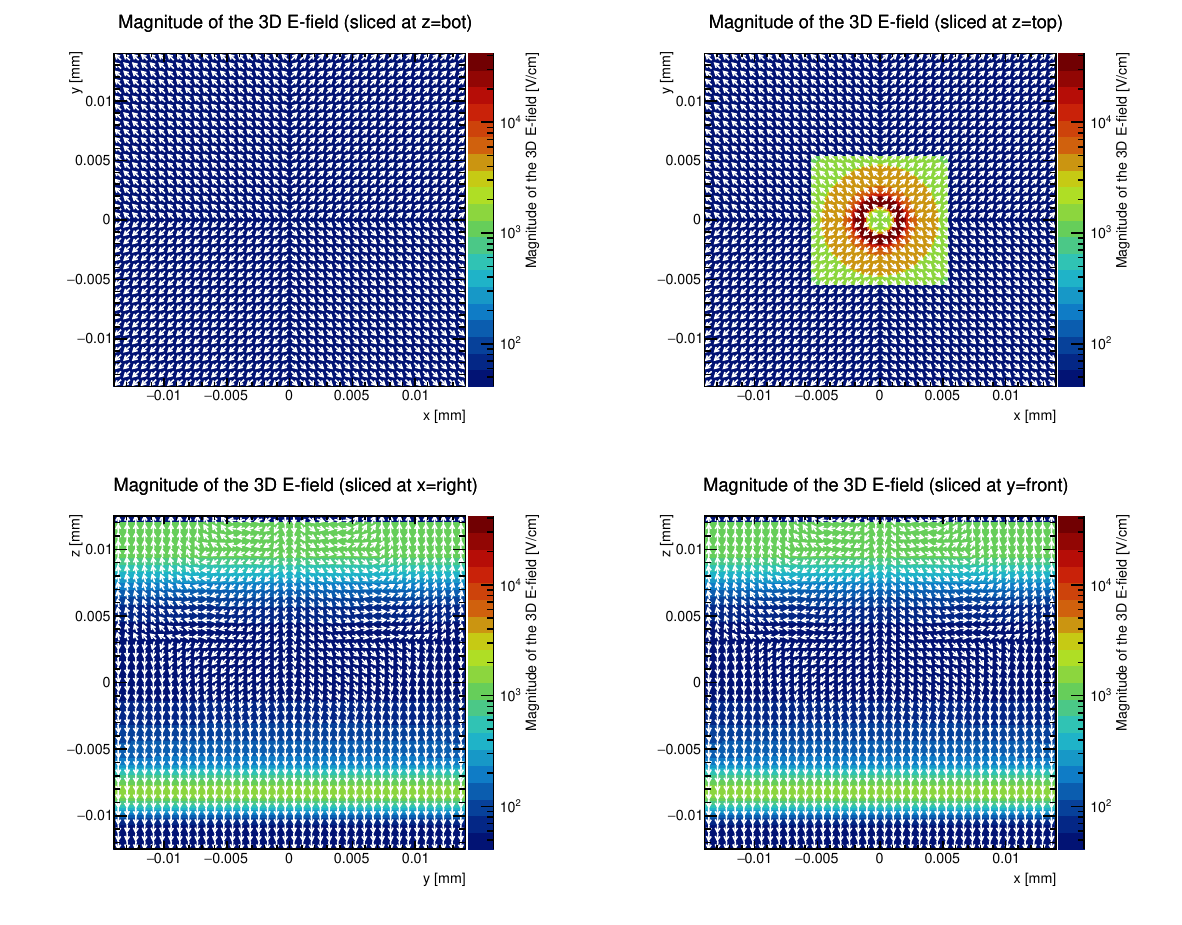}\end{overpic}
\caption{The 3D EFF magnitude at the INVESTIGATOR pixel faces overlaid with the field lines (clockwise from top left: the bottom face, top face, $y$ face and $x$ face).}
\label{fig:effmag2}
\end{figure}

The results shown here are given after the full iterative procedure described in section~\ref{sec:reveng}.
The resulting EFF magnitude of the INVESTIGATOR sensor is seen in figures~\ref{fig:effmag1} and~\ref{fig:effmag2}.
These plots should be compared with figures~4 and~3 of~\cite{DANNHEIM2020163784}.
It can be seen that while the EFF magnitude shape is not identical to the TCAD field map magnitude shape, it captures all important features.
The integral \IntEdl with $\vec{\ell}$ running along the negative $z$-axis direction at the top 25~\um of the sensor in its center results in a bias voltage of -6.08~V.
The analytical expression of the EFF is saved in~\cite{gitEField} in the form of a ROOT TFormula string, along with the code producing it.

\section{Sensor performance with the effective field function}
\label{sec:performance}
The \allpix results shown in this section are using the EFF discussed in section~\ref{sec:elfieldalpide}.
The \allpixEFF results are compared with the ones of the \allpixTCAD and the beam-test data from~\cite{DANNHEIM2020163784}.
The comparison is done at the level of a pixels-cluster in terms of the cluster's charge, size, position residual and efficiency.
The clustering of fired pixels (pixels with charge above threshold) follow a simple algorithm, where all adjacent fired pixels (in both $x$ and $y$ directions) are added together to form a cluster.
In this simple algorithm, no pixel sharing is allowed between two clusters.
The same algorithm is used also in~\cite{DANNHEIM2020163784}.
The cluster size is simply the number of pixels associated with it.
Likewise, the cluster charge is the sum of all pixels' charges associated with it.
The cluster position in $x-y$ is determined, as done in~\cite{DANNHEIM2020163784}, using a variation of the $\eta$ algorithm~\cite{BELAU1983253} to take into account non-linear charge sharing.
This can be compared with the truth position of the incoming $\pi^+$ beam particles to obtain the residuals.
Finally, the detection efficiency can be defined in terms of the fraction of incident primary particles ($\pi^+$) that can be matched to a reconstructed cluster from all primary particles penetrating the detector.
A successful matching is achieved when the maximum distance between the cluster position and the true incident particle position is $100$~\um.
For the comparison, we use a range of thresholds, between 40~e and 700~e, where the nominal threshold is of 120~e.
The number of primary $\pi^+$ beam particles used for the \allpixEFF results below is 20,000 for the nominal as well as the other thresholds.
This gives a low enough statistical error which allows to clearly see the main trends.

Whenever the uncertainties are available in the \allpixTCAD results of~\cite{DANNHEIM2020163784}, we use those.
Whenever the errors are impossible to read from the main distributions/graphs in~\cite{DANNHEIM2020163784}, we take these from the ratio panels\footnote{We note that while the uncertainties in the main panels of~\cite{DANNHEIM2020163784} may appear asymmetric, the uncertainty in the ratio is symmetrised.}.
Therefore, the errors in the results shown below represent the statistical uncertainty of our \allpixEFF simulation and the overall uncertainty from~\cite{DANNHEIM2020163784}.
We do not consider systematic uncertainties for our \allpixEFF results.
However, these should be identical to those of \allpixTCAD since the methods are identical and only the input field map is different.
Therefore, we take the $\delta_{\rm T}$ component of equation~\ref{eq:pull} with a factor of $\sqrt{2}$.
This procedure is valid as long as one excludes systematic variations of the field function itself, a study which we leave for future work.

\begin{figure}[!ht]
\centering
\begin{overpic}[width=0.49\textwidth]{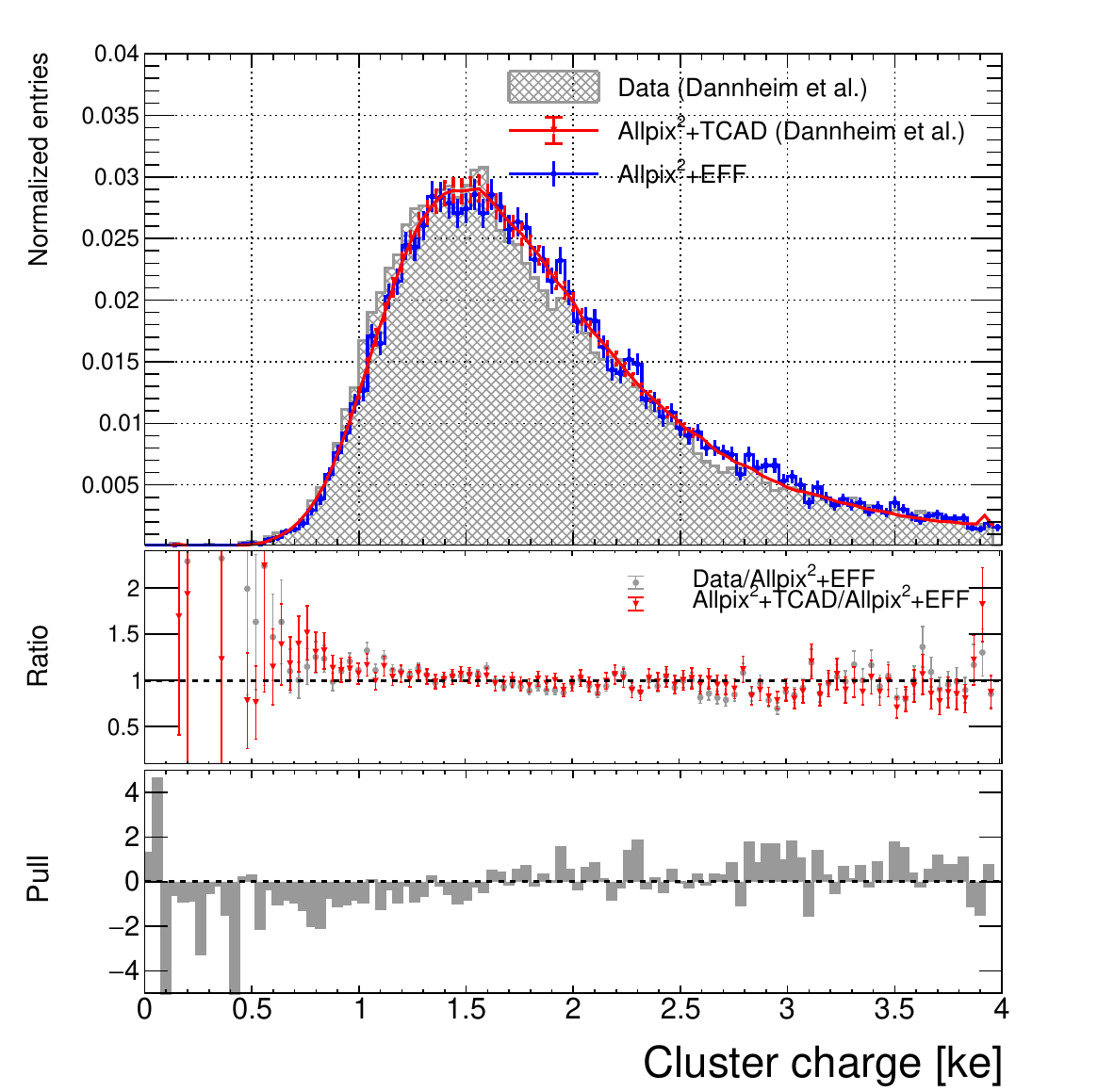}\end{overpic}
\begin{overpic}[width=0.49\textwidth]{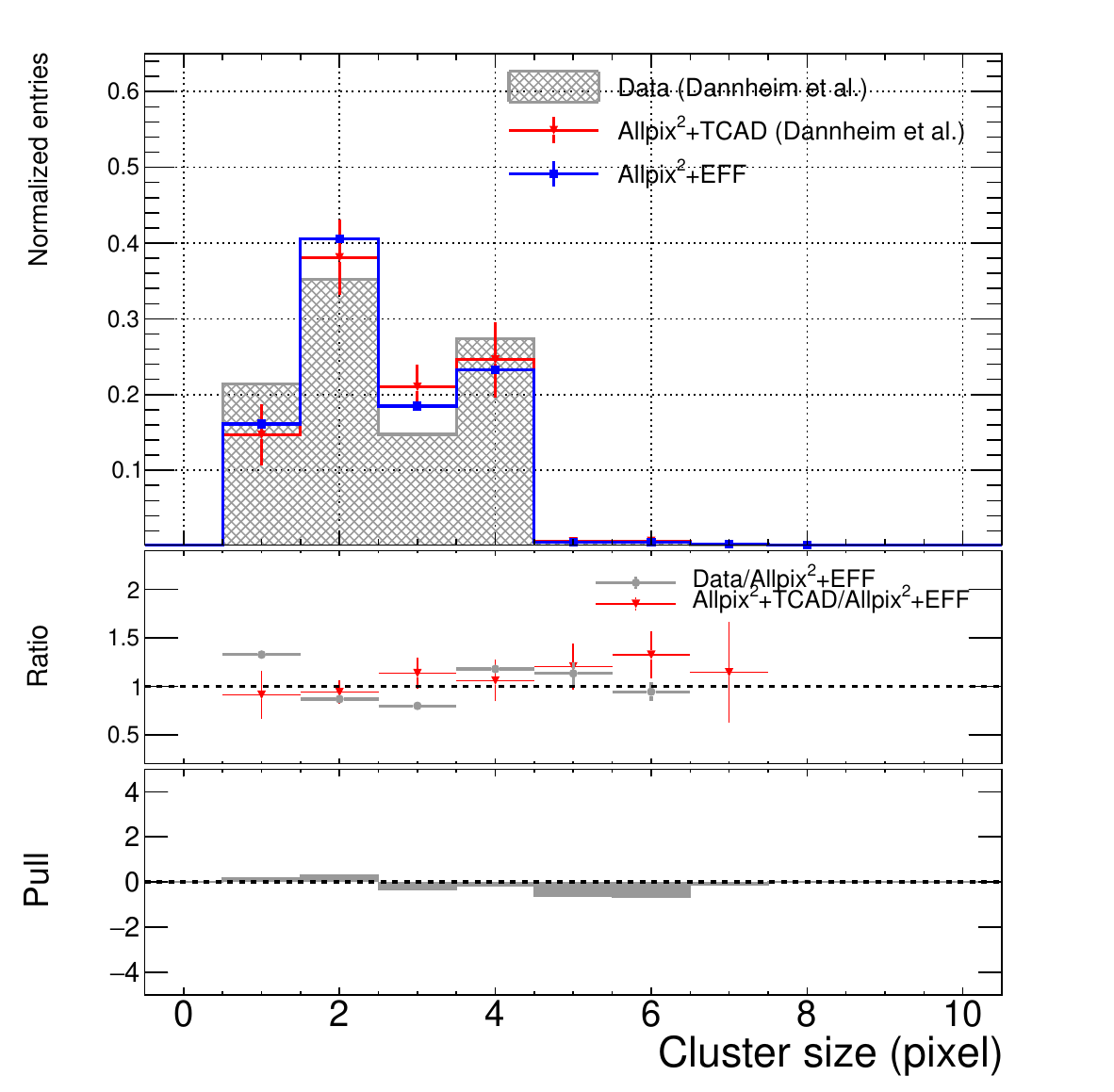}\end{overpic}
\caption{The comparison of the cluster charge (left) and cluster size (right) distributions between our \allpixEFF simulation and the \allpixTCAD simulation (and the beam-test data) from~\cite{DANNHEIM2020163784}.
The middle panels show the ratio of the \allpixTCAD or data results to the \allpixEFF results.
The bottom panels show the pull distributions.
}
\label{fig:charge_size}
\end{figure}

\begin{figure}[!ht]
\centering
\begin{overpic}[width=0.49\textwidth]{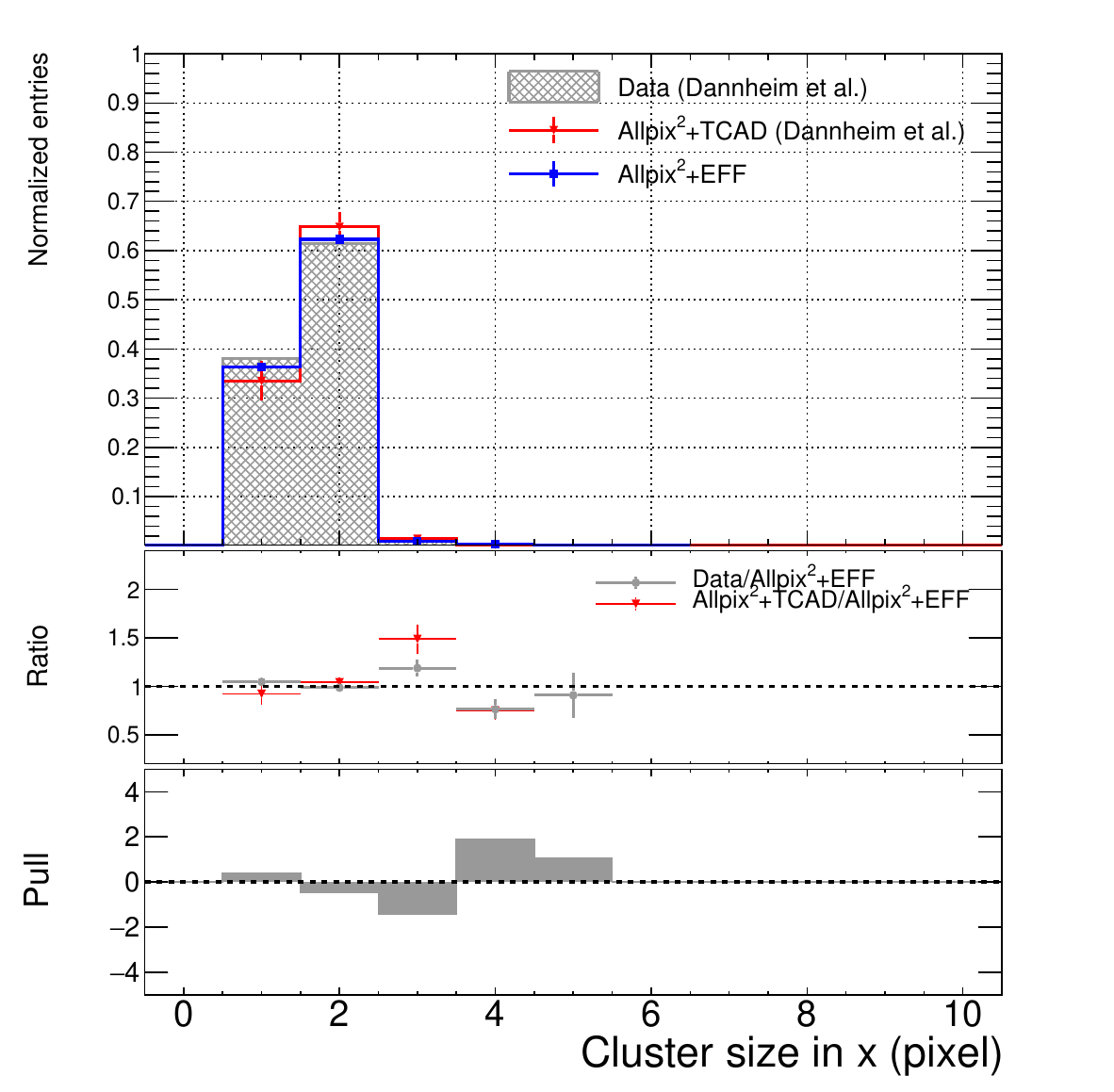}\end{overpic}
\begin{overpic}[width=0.49\textwidth]{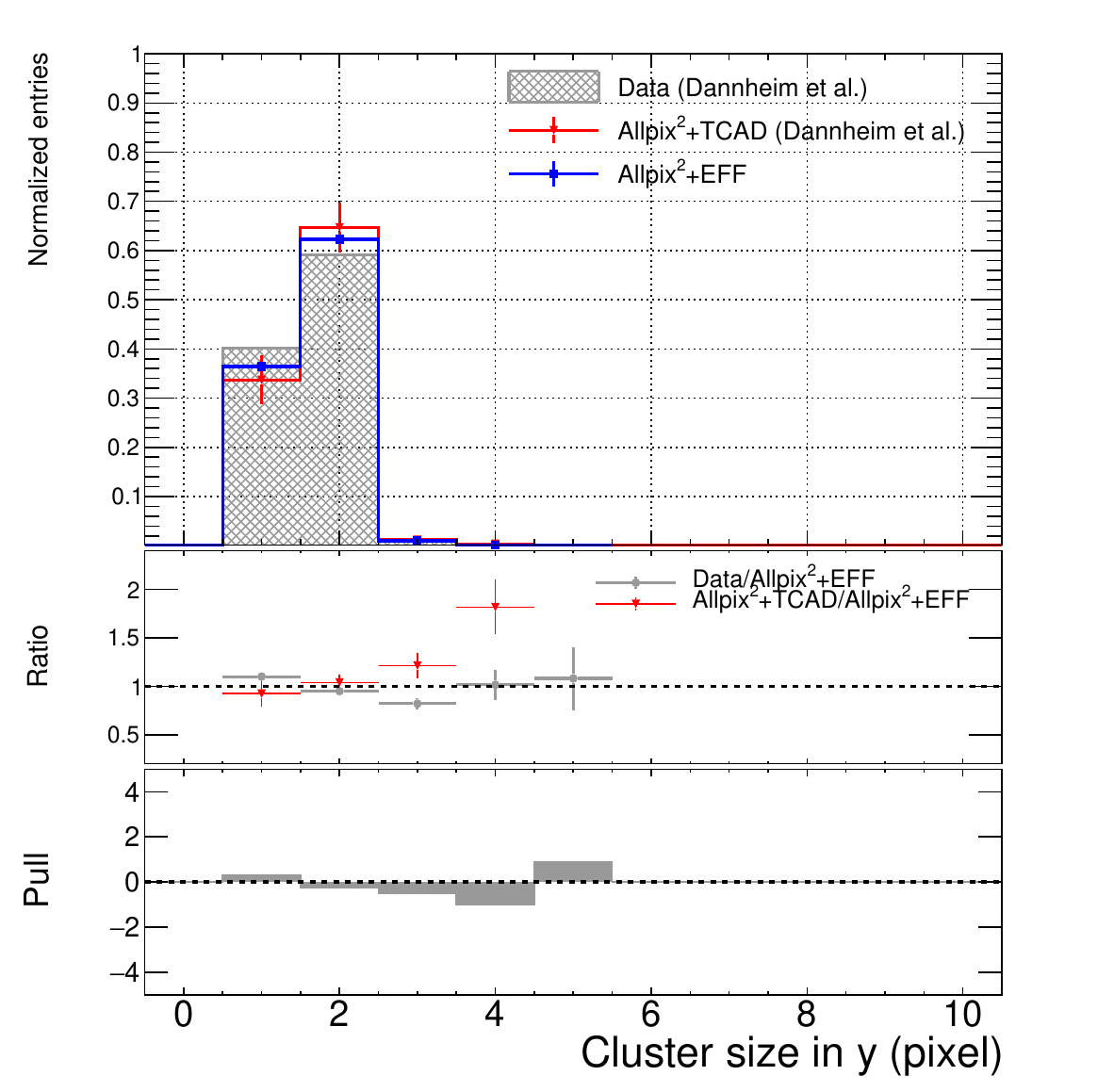}\end{overpic}
\caption{The comparison of the cluster size in $x$ (left) and in $y$ (right) distributions between our \allpixEFF simulation and the \allpixTCAD simulation (and the beam-test data) from~\cite{DANNHEIM2020163784}.
The middle panels show the ratio of the \allpixTCAD or data results to the \allpixEFF results.
The bottom panels show the pull distributions.
}
\label{fig:size_xy}
\end{figure}

\begin{figure}[!ht]
\centering
\begin{overpic}[width=0.65\textwidth]{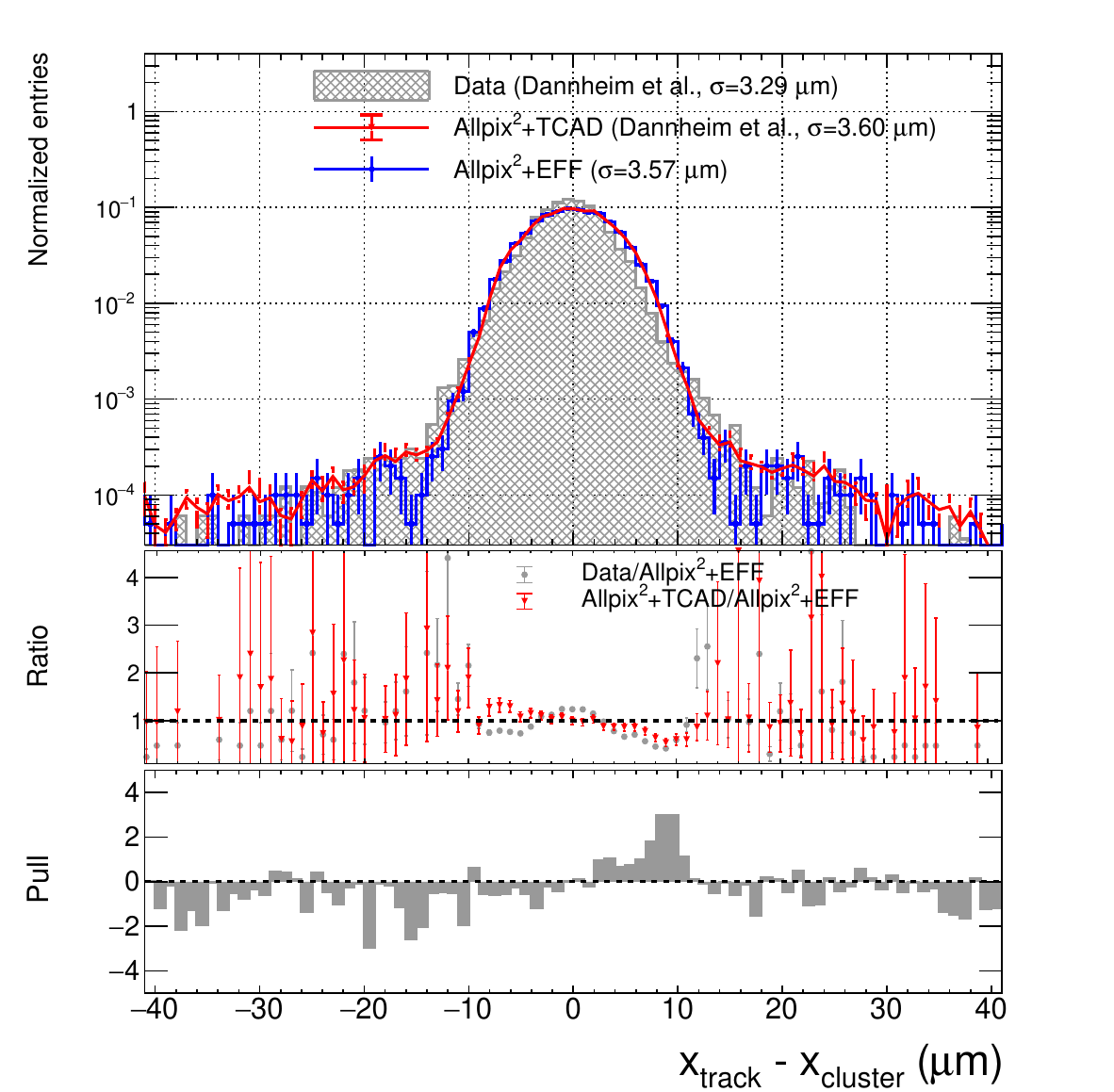}\end{overpic}
\caption{The comparison of the cluster position residuals distribution in $x$ between our \allpixEFF simulation and the \allpixTCAD simulation (and the beam-test data) from~\cite{DANNHEIM2020163784}.
The middle panel show the ratio of the \allpixTCAD or data results to the \allpixEFF result.
The bottom panels show the pull distributions.
The three resolutions reported are calculated in the same way discussed in~\cite{DANNHEIM2020163784}.
}
\label{fig:residuals}
\end{figure}

\begin{figure}[!ht]
\centering
\begin{overpic}[width=0.65\textwidth]{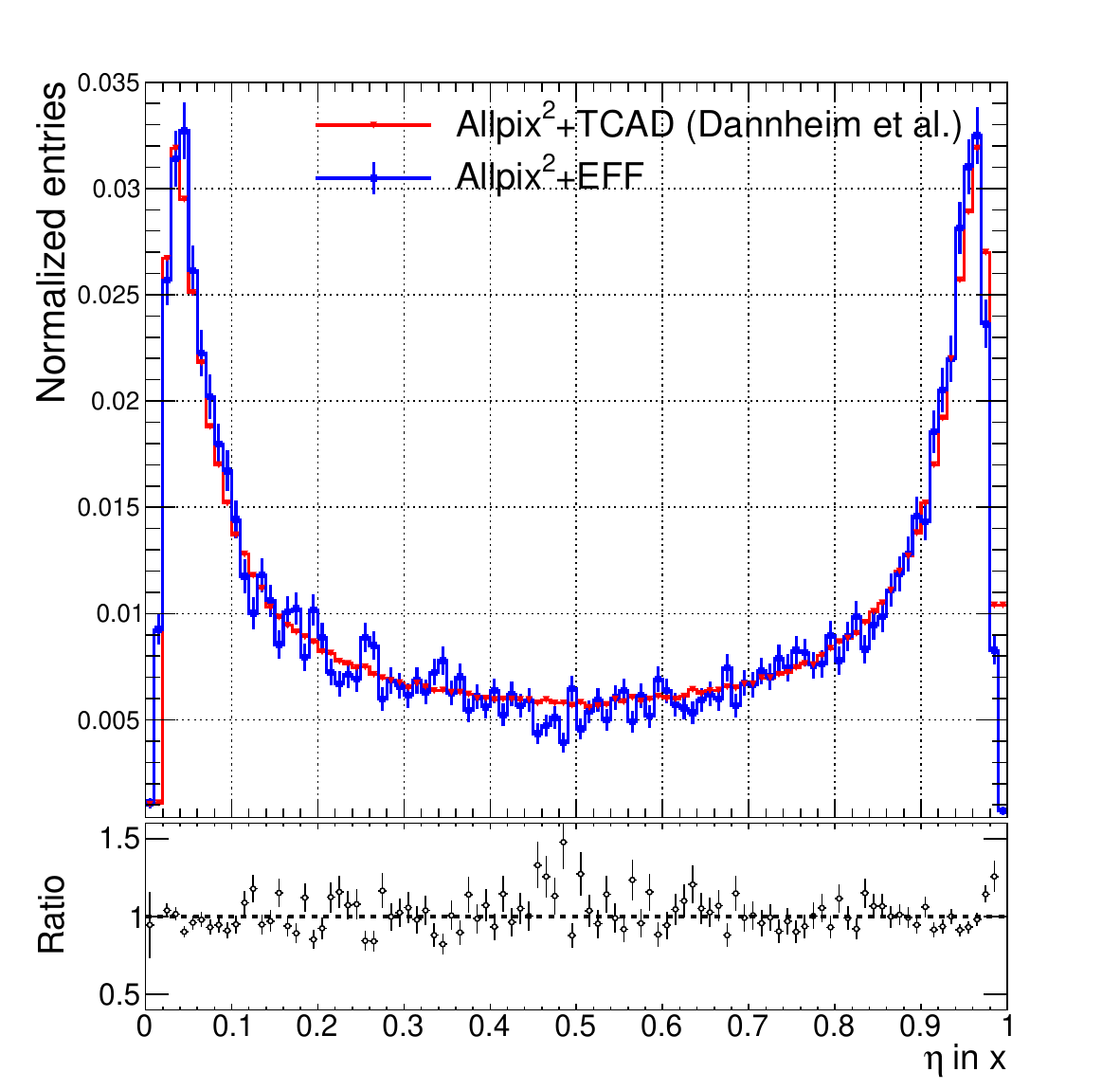}\end{overpic}
\caption{The comparison of the $\eta$ (in $x$) distributions between our \allpixEFF and the \allpixTCAD simulations from~\cite{DANNHEIM2020163784} for a threshold of 40~e.
The bottom panel show the ratio.
The uncertainties are statistical only for the \allpixEFF distribution alone. The \allpixTCAD distribution has no uncertainties in~\cite{DANNHEIM2020163784} and the pull is not shown.
}
\label{fig:eta}
\end{figure}

The cluster charge and size distributions for the nominal threshold are shown in figure~\ref{fig:charge_size}.
The MPV resulting from a fit of the \allpixEFF charge distribution to a convolution of a Landau and a Gaussian probability distribution functions is $1.483\pm0.004$~ke (with $\sigma_{\rm Landau}=0.244\pm0.007$~ke and $\sigma_{\rm Gaus}=0.209\pm0.004$~ke), whereas the MPV result for \allpixTCAD and data is quoted at $1.42$~ke (without uncertainty).
The Gaussian width is $\sigma_{\rm Gaus}=0.21$~ke for data and $\sigma_{\rm Gaus}=0.22$~ke for \allpixTCAD (both quoted without uncertainties).
The MPV and $\sigma_{\rm Gaus}$ values are therefore compatible within $<5\%$.
A good agreement is also seen in the two ratio plots and pull plots, which both comply with the stopping requirements listed above.
The weighted average pull values are $0.60$ and $0.26$, respectively.
The distributions of the cluster size in $x$ and $y$, and the residuals in $x$ are checked as well for the nominal threshold in figures~\ref{fig:size_xy} and~\ref{fig:residuals}.
A similar conclusion about the good agreement can be drawn in these cases as well, where the weighted average pull values are $0.45$, $0.27$ and $0.69$, respectively.
Using figure~\ref{fig:residuals} and following the procedure given in~\cite{DANNHEIM2020163784} to calculate the resolution with the $\eta$ algorithm~\cite{BELAU1983253}, we find that it is $\sigma_x=3.57$~\um for \allpixEFF at a threshold of 120~e, compared with $3.60\pm  0.01({\rm stat})^{+0.24}_{-0.13}({\rm syst})$~\um for \allpixTCAD and $3.29\pm 0.02$ for data.
As mentioned earlier, since the procedures are identical, the systematic uncertainty on our EFF resolution value should be similar to that of \allpixTCAD.
Therefore, our result is compatible not only with the \allpixTCAD simulation but also with the data.
A good agreement is seen in~\ref{fig:eta} for the $\eta$ (in $x$) distribution between the two simulations, but since the source~\cite{DANNHEIM2020163784} has no uncertainties, it is only used for the  residuals rather than as another measure of compatibility.

\begin{figure}[!ht]
\centering
\begin{overpic}[width=0.49\textwidth]{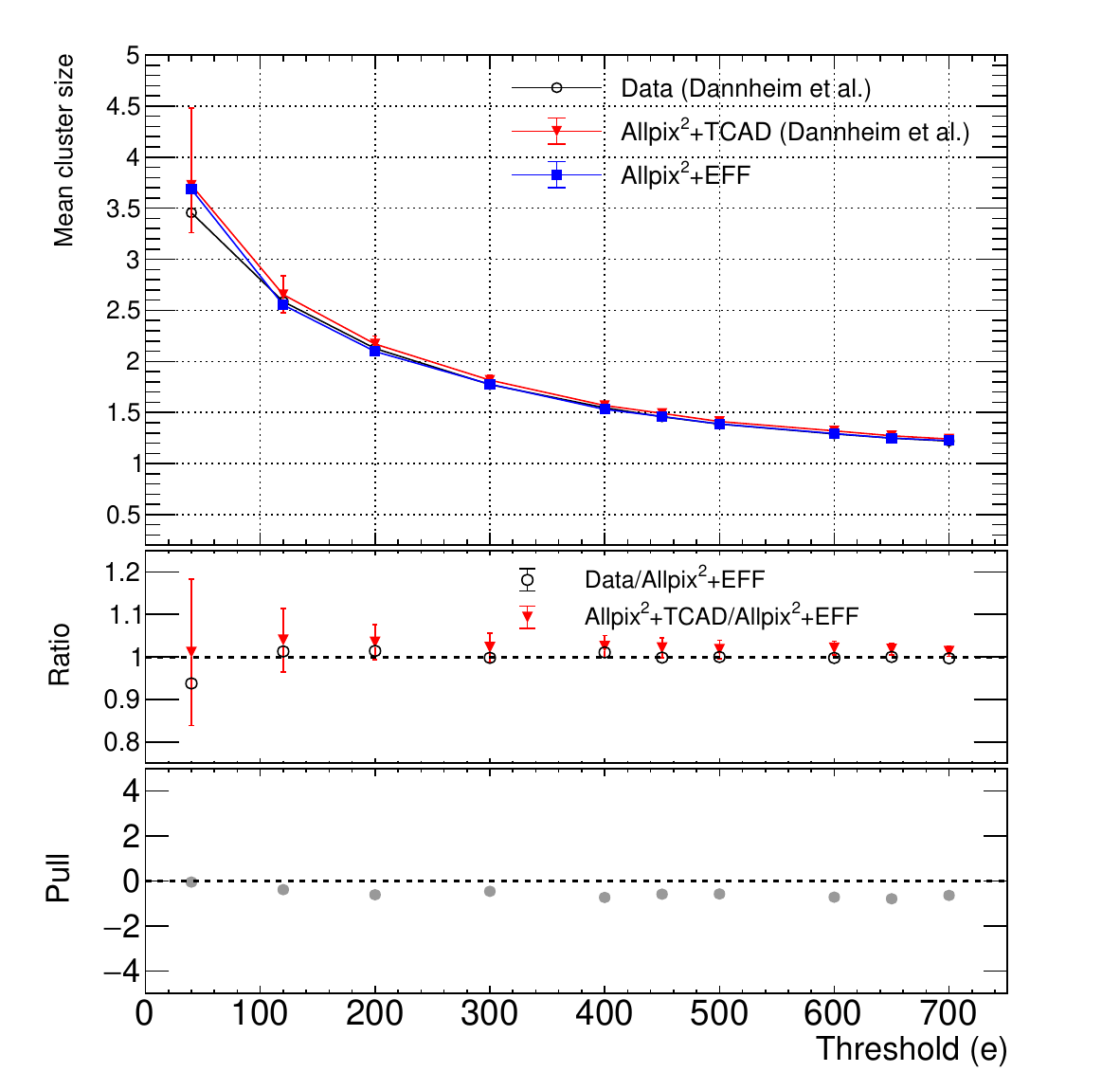}\end{overpic}
\begin{overpic}[width=0.49\textwidth]{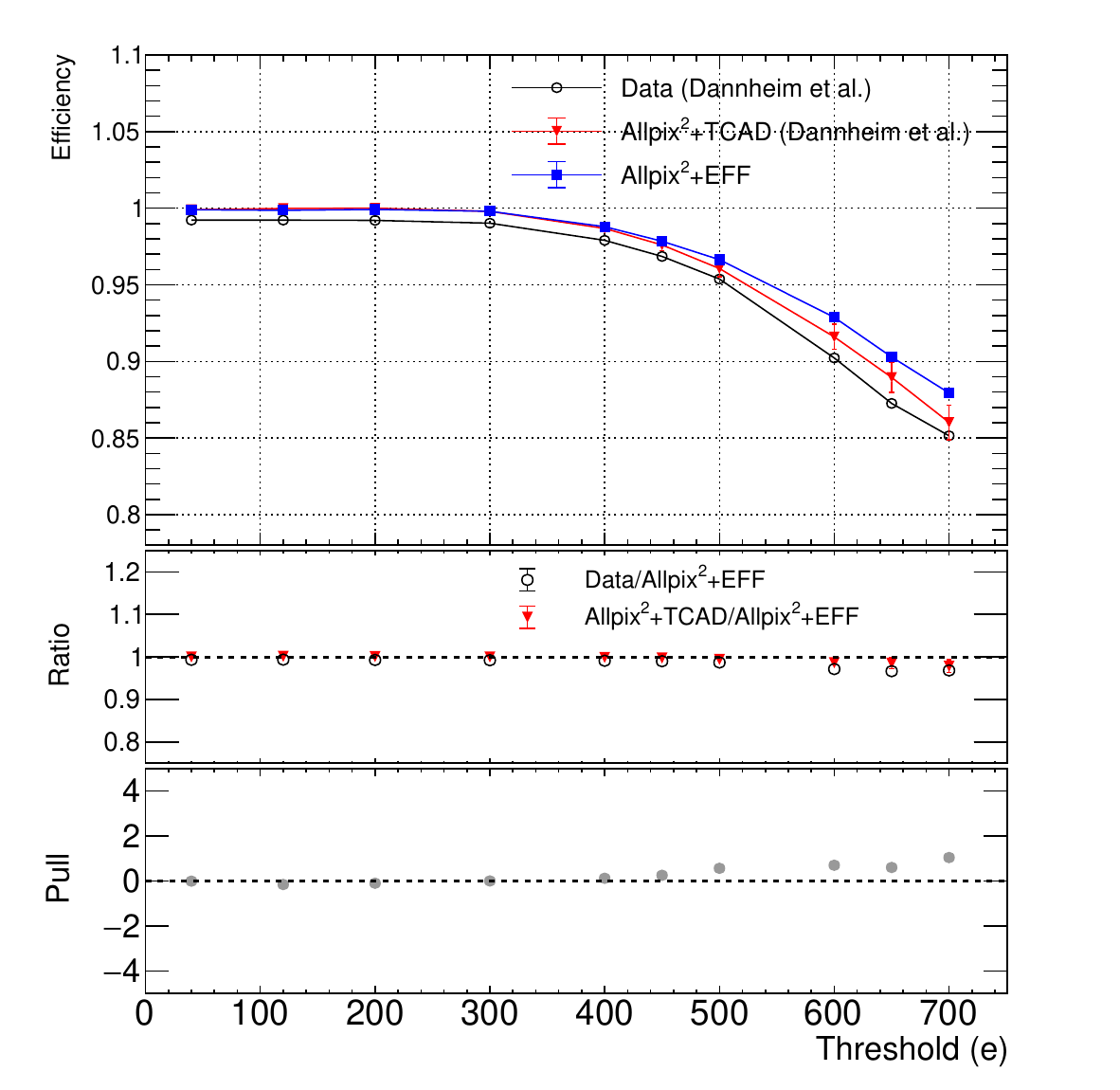}\end{overpic}
\caption{Left: the comparison of the mean cluster size behavior for different thresholds between our \allpixEFF simulation and the \allpixTCAD simulation (and the beam-test data) from~\cite{DANNHEIM2020163784}.
Right: the same comparison for the detection efficiency. 
The middle panels show the ratio of the \allpixTCAD or data results to the \allpixEFF results.
The bottom panels show the pull values.
}
\label{fig:comparisonThreshold}
\end{figure}

\begin{figure}[!ht]
\centering
\begin{overpic}[width=0.49\textwidth]{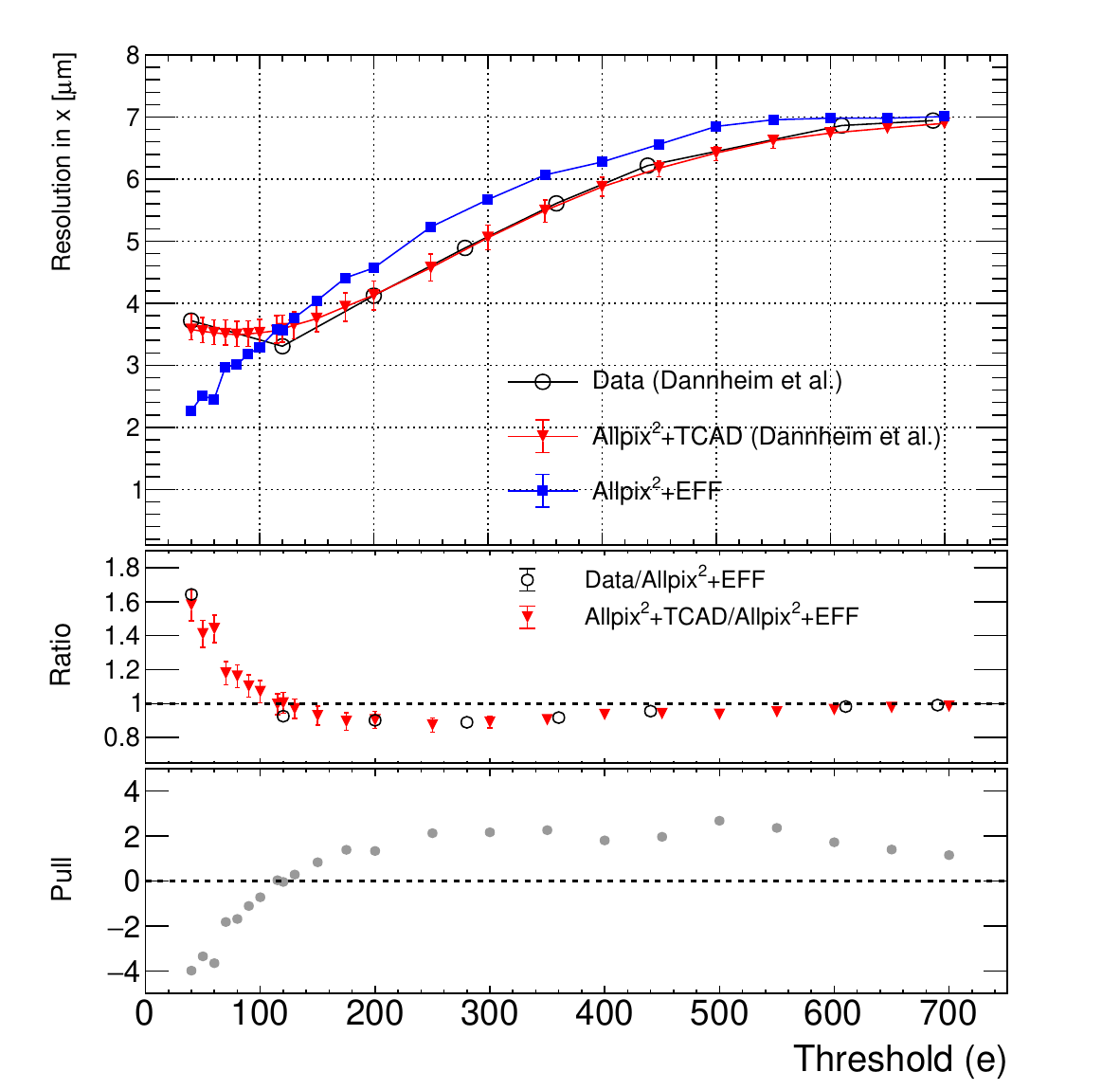}\end{overpic}
\begin{overpic}[width=0.49\textwidth]{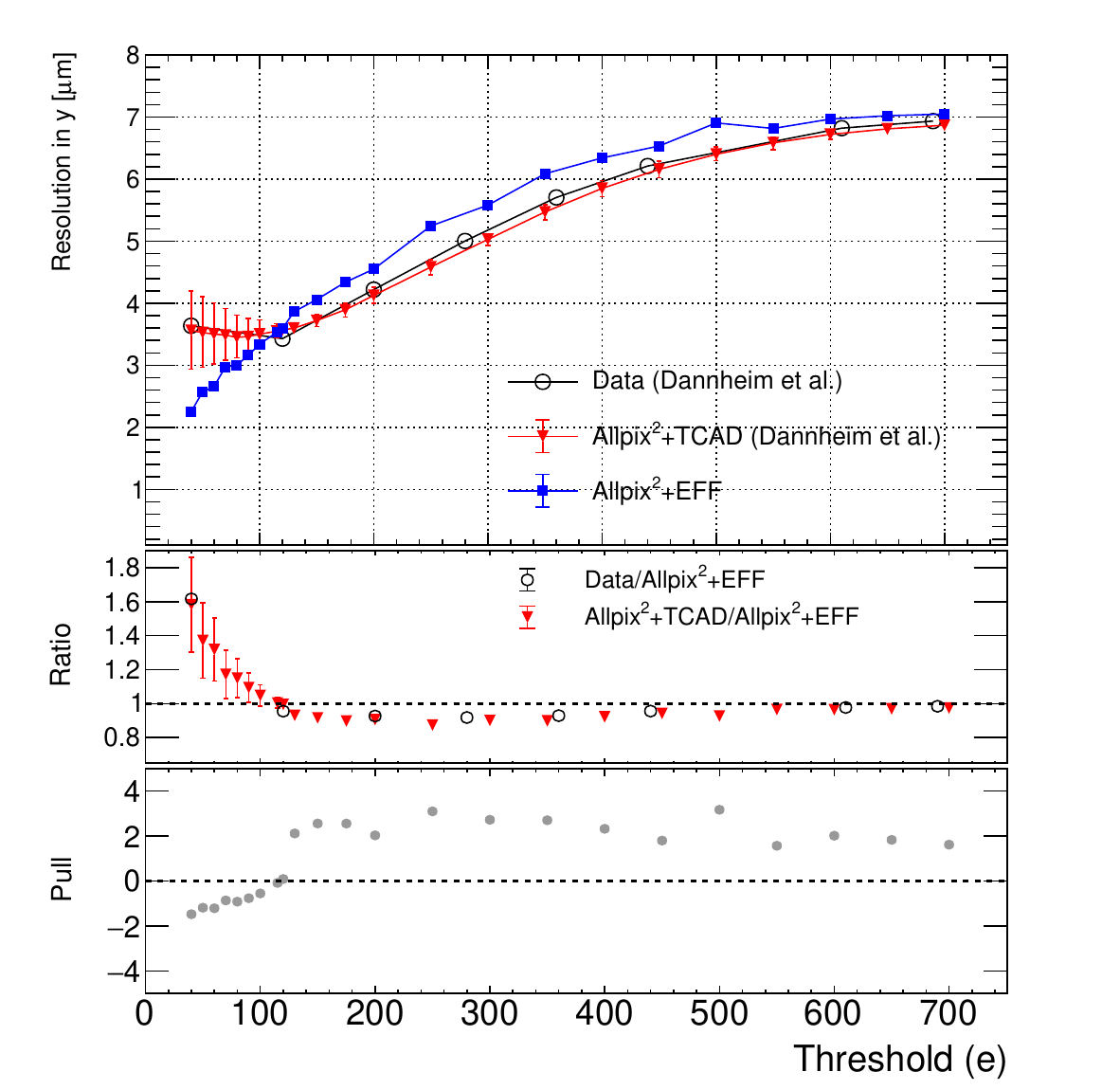}\end{overpic}
\caption{
Left: the comparison of the resolution in $x$ behavior for different thresholds between our \allpixEFF simulation and the \allpixTCAD simulation (and the beam-test data) from~\cite{DANNHEIM2020163784}.
Right: the same comparison for the resolution in $y$.
The middle panels show the ratio of the \allpixTCAD or data results to the \allpixEFF results.
The bottom panels show the pull values.
}
\label{fig:comparisonResThreshold}
\end{figure}

Finally, we also see a good agreement when scanning the thresholds in the range mentioned above and plotting the mean cluster size, the detection efficiency and the spatial resolution in $x$ and $y$ vs the threshold.
These results are shown in figures~\ref{fig:comparisonThreshold} and~\ref{fig:comparisonResThreshold}.
The largest discrepancy between the two results is seen in the spatial resolution at low thresholds, $\lesssim 100$~e.
We note that in most use-cases the sensors are not operated with thresholds below 100~e.
We also see that the ratio between the two results is compatible within $<10\%$ above $\simeq 100$~e.
The pull graphs are showing a consistent incompatibility at the level of $+2$ in that range.
The maximum discrepancy at the range above 100~e is nevertheless at the level of $\sim 0.5$~\um.
This difference is $<10\%$ of the best resolution measured for the INVESTIGATOR sensor~\cite{ALPIDE3} and it is negligible for all practical purposes.

This small difference in resolution cannot be explained by small differences in the cluster size distributions since we see that the mean size of the full cluster in figure~\ref{fig:comparisonThreshold} (left) is very similar between the two simulations for all thresholds tested.
A possible explanation may be the slightly different modelling of the charge sharing between the two fields (EFF and TCAD), hinted in the edge of the $\eta$ distribution at 40~e in figure~\ref{fig:eta}.
However, since these distributions (in $x$ and $y$) are not included in~\cite{DANNHEIM2020163784} for thresholds other than 40~e, we cannot confirm this hypothesis.

Finally, we note that the EFF is valid under the conditions for which it is derived, using a bias voltage of $-6$~V.
While it may be applicable also for other voltage values (by changing its overall normalization), it will be invalid for zero bias.

\section{Conclusions and outlook}
\label{sec:concl}
We show that a carefully constructed EFF may replace a highly detailed TCAD field map in the simulation chain of semiconductor sensors in case the latter is not available. 
This is shown for the highly non-linear and non-trivial field of the INVESTIGATOR sensor, where the TCAD field map is not available publicly.
Our EFF reproduces the same performance as shown in~\cite{DANNHEIM2020163784} to a very good extent for several observable quantities in a wide range of values and operable configurations.
The compatibility between our \allpixEFF and the \allpixTCAD (and data) results is particularly good at the nominal threshold of 120~e considered in~\cite{DANNHEIM2020163784} and in general, throughout the sensors' typical range of operation, above $\sim$100~e.
We point out that, while our EFF (available in~\cite{gitEField}) can be already used now, it may also be further optimized to reach an even better agreement targeting the discrepancy of $\sim 0.5$~\um seen in the $x$ and $y$ resolution for thresholds in the range of $\sim$150-600~e (or $\sim 1.5$~\um below 80~e).
In that sense, the decision where to stop the optimization process is purely driven by practical considerations, which depend on the application.

Our work allows the many ALPIDE sensor users worldwide, who do not have access to the TCAD field map, to use our EFF as an important input for their detector simulation.
In fact, the tracking detector simulation of the LUXE experiment~\cite{LUXECDR}, which is to be built from the production version of the ALPIDE sensors, is already using our EFF.
The ALPIDE sensor has slightly different dimensions compared to the INVESTIGATOR sensor used in~\cite{DANNHEIM2020163784}.
The EFF can be therefore naturally re-scaled by changing the dimensions of the pixel since its components are defined with respect to the pixel dimensions in 3D.
Effectively only the pitch needs to very slightly change, where the change in thickness has no impact on the field since it is only defined within the epitaxial layer.
The re-scaling procedure is similar to the procedures used in the ALICE experiment simulation of the ALPIDE sensors~\cite{MIKOPAPER}.

The procedures discussed in this work can be regarded as general as long as there is at least (i) a very basic knowledge on the field's shape and (ii) reference data for the comparison of the sensor performance with simulation, from beam-tests, radioactive sources, cosmic rays, etc.
Finally, we note that although we use the \allpix software extensively for the derivation and validation of our EFF, the resulting function can be interfaced also with other software like~\cite{MIKOPAPER} and it is not linked specifically to \allpix.

\clearpage
\newpage
\section*{Acknowledgments}
We wish to thank the colleagues from the ALICE ITS project for useful discussions and help related to the ALPIDE sensors.
We would like to especially thank Luciano Musa, Gianluca Aglieri Rinella, Antonello Di Mauro, Magnus Mager, Corrado Gargiulo, Felix Reidt, Ivan Ravasenga, and Ruben Shahoyan.
We also wish to thank Simon Spannagel and Paul Sch\"utze for the detailed and dedicated help and for the useful discussions about \allpix and especially the work in~\cite{DANNHEIM2020163784}.\\

\noindent This work is supported by a research grant from the Estate of Dr. Moshe Gl\"{u}ck, the Minerva foundation with funding from the Federal German Ministry for Education and Research, the ISRAEL SCIENCE FOUNDATION (grant No. 708/20), the Anna and Maurice Boukstein Career Development Chair, the Benoziyo Endowment Fund for the Advancement of Science, the Estate of Emile Mimran and the Estate of Betty Weneser.

\appendix

\bibliographystyle{elsarticle-num}
\bibliography{Efield_MAPS}
\end{document}